\documentclass[a4paper,11pt]{article}
\pdfoutput=1
\usepackage{jheppub}
\usepackage{amsmath,amssymb,latexsym}
\usepackage{subcaption}
\usepackage{braket}
\usepackage{bm}
\usepackage{hepunits}
\usepackage{graphicx}
\usepackage[svgnames]{xcolor}

\usepackage{mathtools}

\graphicspath{{.}{figure}}

\newcommand{\dif}{\mathrm{d}}

\newcommand{\w}{\omega}

\allowdisplaybreaks

\DeclareMathOperator{\Res}{Res}

\title{Intersection theory, relative cohomology and the Feynman parametrization}
\author{Mingming Lu,}
\author{Ziwen Wang,}
\author{Li Lin Yang}
\affiliation{Zhejiang Institute of Modern Physics, School of Physics, Zhejiang University, Hangzhou 310027, China}

\emailAdd{mingming.lu@zju.edu.cn}
\emailAdd{zwenwang@zju.edu.cn}
\emailAdd{yanglilin@zju.edu.cn}

\abstract{We present a novel approach for loop integral reduction in the Feynman parametrization using intersection theory and relative cohomology. In this framework, Feynman integrals correspond to boundary-supported differential forms in the language of relative cohomology. The integral reduction can then be achieved by computing intersection numbers. We apply our method in several examples to demonstrate its correctness, and discuss the subtleties in certain degenerate limits.}

\begin{document}

\maketitle

\section{Introduction}

Feynman integrals are essential ingredients of perturbative quantum field theories. For cutting-edge phenomenological applications, it is of high importance to develop efficient methods for their calculation. A complicated multi-loop scattering amplitude may involve a gigantic amount of Feynman integrals. A first step to evaluate the amplitude is to employ the linear relations among the Feynman integrals, and reduce them to a finite set of master integrals (MIs). This not only significantly reduces the number of Feynman integrals to calculate, but is also a crucial step in the method of differential equations \cite{Kotikov:1990kg, Kotikov:1991pm, Remiddi:1997ny, Gehrmann:1999as}.

Currently, the standard method for multi-loop integral reduction is the Laporta algorithm \cite{Laporta:2000dc, Laporta:2000dsw} that solves the integration-by-parts (IBP) identities \cite{Tkachov:1981wb, Chetyrkin:1981qh} of Feynman integrals. There are already numerous software tools implementing the IBP reduction, including \texttt{AIR}~\cite{Anastasiou:2004vj}, \texttt{FIRE}~\cite{Smirnov:2023yhb}, \texttt{LiteRed}~\cite{Lee:2013mka}, \texttt{Reduze}~\cite{vonManteuffel:2012np}, \texttt{Kira}~\cite{Klappert:2020nbg} and \texttt{AmpRed}~\cite{Chen:2019mqc, Chen:2019fzm, Chen:2020wsh, Chen:2024xwt}. There are also recent efforts to generate a smaller set of IBP identities, such as \texttt{NeatIBP}~\cite{Wu:2023upw} and \texttt{Blade}~\cite{Guan:2024byi}. This small set of identities potentially leads to a speed up of the reduction procedure.

The intersection theory \cite{zbMATH00713739, zbMATH01270294, zbMATH02112802, zbMATH06267077, zbMATH06447521, zbMATH06502598, zbMATH06454357, zbMATH07531013, zbMATH07527773, zbMATH07733418} provides an intriguing new perspective to formulate IBP relations. It was initially developed in algebraic geometry to analyze general (Aomoto-Gelfand) hypergeometric functions, and has proven applicable across various fields in mathematics and theoretical physics. By elucidating the vector space structure of so-called \textit{period integrals}, it provides an alternative framework for computing Feynman integrals and scattering amplitudes~\cite{Mizera:2017rqa, Mastrolia:2018uzb}. A family of period integrals is generated by a specific multivalued function called \textit{twist}, along with a family of meromorphic differential forms. Such an integral family exhibits a finite-dimensional vector space structure, which can be articulated within the framework of twisted de Rham cohomology. The vectors in such a cohomology group represent equivalence classes of integrals.

The twisted cohomology group can be endowed with an inner product structure known as \textit{intersection numbers}. They can be used to facilitate the decomposition of a vector into the linear combination of a basis. When applied to Feynman integral reduction \cite{Mastrolia:2018uzb, Frellesvig:2019kgj, Frellesvig:2019uqt, Mizera:2019vvs, Mizera:2020wdt, Frellesvig:2020qot, Caron-Huot:2021xqj, Caron-Huot:2021iev, Chestnov:2022alh, Fontana:2023amt, Brunello:2023rpq}, this decomposition is completely equivalent to IBP reduction, and the basis simply corresponds to a set of MIs. Besides the reduction of Feynman integrals, this framework is applicable to a wider range of mathematical systems, including Aomoto-Gelfand hypergeometric functions, Euler-Mellin integrals and Gelfand-Kapranov-Zelevinsky systems~\cite{zbMATH04188193, zbMATH00007653, zbMATH05821564}. Furthermore, it has demonstrated great capability in various areas of theoretical physics. For example, it has been applied to the construction of canonical bases for Feynman integrals and the derivation of canonical differential equations~\cite{Chen:2020uyk, Chen:2022lzr, Chen:2023kgw, Chen:2024ovh}, to the computation of correlation functions in lattice gauge theories~\cite{Weinzierl:2020nhw, Gasparotto:2022mmp, Gasparotto:2023roh}, to the solution of quantum mechanical problems~\cite{Cacciatori:2022mbi}, to gravitational wave physics in classical general relativity \cite{Brunello:2024ibk, Frellesvig:2024swj}, to Fourier calculus~\cite{Brunello:2023fef}, to string theory \cite{Bhardwaj:2023vvm}, and to the study of cosmological correlators~\cite{De:2023xue, Benincasa:2024ptf}.

Given the wide applicability of intersection theory, the calculation of intersection numbers has recently become an important topic under investigation. For the special case where the differential forms are logarithmic, the computation of intersection numbers is straightforward~\cite{zbMATH01270294, zbMATH07733418}. For general multivariate differential forms, the most established method is to perform the computation variable-by-variable \cite{Frellesvig:2019uqt, Frellesvig:2020qot}, although method employing multivariate higher-order partial differential equations is also available \cite{Chestnov:2022xsy}. In the variable-by-variable approach, the intersection numbers are calculated recursively. For each variable, one needs to compute a sum of residues at the singularities determined by the twist. These singularities are roots of a polynomial, which often involve algebraic extensions. To avoid explicitly working with algebraic extensions, one may employ the global residue theorem to calculate the sum of residues with the help of polynomial division \cite{Weinzierl:2020xyy,Fontana:2023amt, Brunello:2023rpq}. When combined with the idea of companion matrices \cite{Brunello:2024tqf}, this technique has been applied to cutting-edge problems in Feynman integral reduction.

In the context of integral reduction, it is often the case that the singularities of the differential forms are different from those of the twist. To define the intersection numbers, one needs to introduce extra regulators, and take them to zero after performing the integral decomposition. Alternatively, one may employ the concept of relative cohomology \cite{zbMATH07905779, Caron-Huot:2021xqj, Caron-Huot:2021iev, Brunello:2023rpq}, which treats the twisted boundaries (singularities of the twist) and the relative boundaries (singularities of the differential forms) differently. When working with the Baikov representation \cite{Baikov:1996iu, Lee:2010wea} of Feynman integrals, the differential forms are singular when the propagator denominators approach zero. One may introduce boundary-supported $\delta$-forms as duals of Feynman integrals, and use relative cohomology to compute the intersection numbers. This approach avoids the necessity of intermediate regulators, and can substantially improve the computational efficiency.

In this paper, we initiate the application of relative cohomology and intersection theory to the Feynman parametrization of loop integrals. There are a few benefits when working with the Feynman parametrization. The Symanzik polynomials appearing in the Feynman parametrization are usually simpler than the Baikov polynomial. They are homogeneous polynomials of the Feynman parameters, and can be naturally interpreted in a projective space. In the Feynman parametrization, the recursive structure of representations in different sub-sectors of an integral family is manifest. The integrals in all sectors can be represented using the Symanzik polynomials of the top sector, and the sub-sectors are characterized by one or more $\delta$-functions in the integrals. This should be contrasted with the recursive structure of the Baikov representation \cite{Jiang:2023qnl, Jiang:2023oyq}, where one needs to integrate out a couple of variables to arrive at sub-sectors. In the Feynman parametrization, one also does not need to worry about irreducible scalar products (ISPs). The loop integrals with ISPs in the numerator can be easily converted to integrals with Symanzik polynomials in the denominator, which can be treated straightforwardly using intersection theory. Finally, from the above discussions, we observe that the sub-sector integrals can be naturally regarded as boundary-supported differential forms in relative cohomology. We may then perform the integral reduction with intersection theory without introducing extra regulators for relative boundaries.

The paper is organized as follows. In Section~\ref{sec:IntThLP}, we introduce the ingredients of our method, including the Feynman parametrization, the intersection theory and the concept of relative cohomology. In section~\ref{sec:Examples}, we provide several examples to demonstrate the correctness of our method, and point out the subtleties in degenerate limits. We summarize in Section~\ref{sec:Summary}.

\section{Intersection theory for the Feynman parametrization}\label{sec:IntThLP}

\subsection{Feynman parametrization}

An $L$-loop Feynman integral is defined by
\begin{equation}
    I(\nu_1,\cdots,\nu_n) = e^{\epsilon\gamma_E L} \int \frac{\dif^dk_1}{i\pi^{d/2}} \cdots \frac{\dif^dk_L}{i\pi^{d/2}} \, \frac{1}{D_1^{\nu_1} \cdots D_n^{\nu_n}} \,,
\end{equation}
where $D_i$ are propagator denominators or irreducible scalar products. We will work with a slightly different version of the Feynman parametrization proposed by Lee and Pomeransky (the so-called LP parametrization) \cite{Lee:2013hzt}:
\begin{equation}
    I(\nu_1,\cdots,\nu_n) = e^{\epsilon\gamma_E L} \, \frac{(-1)^\nu \, \Gamma(d/2)}{\Gamma\left((L+1)d/2-\nu\right)} \int_0^\infty \left( \prod_{i} \frac{x_i^{\nu_i-1} \dif x_i}{\Gamma(\nu_i)} \right) \left( \mathcal{U} + \mathcal{W} \right)^{-d/2} \,,
\end{equation}
where $\nu \equiv \sum_j \nu_j$, $\mathcal{U}$ and $\mathcal{W}$ are the so-called Symanzik polynomials. We denote
\begin{equation}
    x_1 D_1 + \cdots + x_n D_n \equiv \sum_{i,j=1}^L  M_{ij} \, k_i \cdot k_j - 2 \sum_{i=1}^L k_i \cdot Q_i - J + i0 \,,
\end{equation}
where $Q_i$ are combinations of external momenta. The two Symanzik polynomials can then be written as
\begin{equation}
    \mathcal{U} = \det(M) \, , \quad
    \mathcal{W} = \det(M) \left[ \sum_{i,j=1}^L M^{-1}_{ij} \, Q_i \cdot Q_j - J - i0 \right] .
    \label{eq:uw_poly}
\end{equation}
From the above expressions, it is clear that $\mathcal{U}$ and $\mathcal{W}$ are homogeneous polynomials of degree $L$ and $L+1$ in the variables $x_i$, respectively. 

Suppose $\mathcal{S}$ is a non-empty subset of $\{1,2,\cdots,n\}$. In the LP parametrization, we can insert
\begin{equation}
    1 = \int_0^\infty \dif\eta \, \delta \left( \eta - \sum_{i \in \mathcal{S}} x_i \right) ,
\end{equation}
and then rescale $x_i \to \eta x_i$. Integrating over $\eta$, we arrive at the standard Feynman parametrization:
\begin{multline}
    I(\nu_1,\cdots,\nu_n) = e^{\epsilon\gamma_E L} \, (-1)^\nu \, \Gamma(\nu-Ld/2)
    \\
    \times \int_0^1 \left( \prod_{i \in \mathcal{S}} \frac{x_i^{\nu_i-1} \dif x_i}{\Gamma(\nu_i)} \right) \delta \left( 1 - \sum_{i \in \mathcal{S}} x_i \right) \int_0^\infty \left( \prod_{j \notin \mathcal{S}} \frac{x_j^{\nu_j-1} \dif x_j}{\Gamma(\nu_j)} \right) \frac{\mathcal{U}^{\nu-(L+1)d/2}}{\mathcal{W}^{\nu-Ld/2}} \,.
\end{multline}

In the above, we have assumed that all $\nu_i > 0$. If some of the $\nu_i$'s are zero, the corresponding propagator denominators $D_i$ are actually absent in the integral. These kind of integrals belong to sub-sectors within the integral family. In this case, we don't need to introduce the $x_i$ variable at all into the Feynman parametrization. Equivalently, we can insert a $\delta(x_i)$ into the integration measure, and get rid of the $\Gamma(\nu_i)$ in the denominator. This is also equivalent to introducing a regulator $D_i^{-\rho}$ in the integrand. After Feynman parametrization, this becomes
\begin{equation}\label{eq:rho_delta}
    \frac{x_i^{\rho - 1}}{\Gamma(\rho)} = \delta(x_i) + \mathcal{O}({\rho^1}) \,.
\end{equation}
By introducing these $\delta(x_i)$, we can express integrals in all sectors of the family using a unified LP polynomial $G \equiv \mathcal{U} + \mathcal{W}$. To ensure mathematical rigor, we emphasize that the Dirac $\delta$-functions employed in this framework constitute Radon measures within the context of Lebesgue integrals. These are related to the concept of relative cohomology to be introduced later. For these measure-theoretic concepts, we refer the readers to Chapter~6 of \cite{rudin1991functional} for a comprehensive introduction.

With the regulators as above, we can also incorporate the situations where some $\nu_i < 0$. The corresponding $D_i$'s are present in the numerator instead of in the denominator, and are usually called irreducible scalar products (ISPs) in the literature. In this case, we can integrate by parts and obtain\footnote{Alternatively, one may perform a formal expansion in terms of derivatives of the $\delta$-function, as discussed in \cite{Brunello:2023rpq}.}
\begin{align}
    \int_{0}^{\infty} \dif x_i \, \frac{x_i^{\rho+\nu_i-1}}{\Gamma(\rho+\nu_i)} \, G^{-d/2} &= \int_{0}^{\infty} \dif x_i \, \frac{ x_i^{\rho-1}}{\Gamma(\rho)} \left( -\frac{\partial}{\partial x_i} \right)^{-\nu_i} G^{-d/2} \nonumber
	\\
	&= \int_{0}^{\infty} \dif x_i \, \delta(x_i) \left( -\frac{\partial}{\partial x_i} \right)^{-\nu_i} G^{-d/2} + \mathcal{O}(\rho^1) \,.
	\label{eq:ibp_for_num}
\end{align}
Therefore, we can again express the integrals using the same $G$ polynomial. Note that the action of the partial derivatives will result in a factor $G^{\nu_i}$ multiplying the usual $G^{-d/2}$. These correspond to integrals in shifted spacetime dimensions, and are often regarded as troublesome in many IBP reduction methods. However, it is a trivial problem in methods based on intersection theory, since integrands with extra factors of $G$ in the denominator are treated automatically in a unified way. 

In the following, we will strip the pre-factors and be concerned with the family of integrals
\begin{align}
    J(\nu_1,\cdots,\nu_n) &= \int_0^\infty \left( \prod_{i,\,\nu_i>0} x_i^{\nu_i-1} \, \dif x_i \right) \left( \prod_{j,\, \nu_j \le 0} \delta(x_j) \dif x_j \left( -\frac{\partial}{\partial x_j} \right)^{-\nu_j} \right) G^{-d/2} \nonumber
	\\
	&= \int_0^\infty \left( \prod_{i,\,\nu_i>0} x_i^{\nu_i-1} \, \dif x_i \right) R_0 \, G_0^{-d/2} \,,
	\label{eq:LP_J}
\end{align}
where
\begin{align}
G_0 \equiv G(x_j \to 0, \forall \nu_j \leq 0) \,, \quad R_0 \equiv G^{d/2} \left( \prod_{j,\, \nu_j \le 0} \left( -\frac{\partial}{\partial x_j} \right)^{-\nu_j} \right) G^{-d/2} \Bigg|_{x_j \to 0, \forall \nu_j \leq 0} \,.
\label{eq:LP_G0}
\end{align}
The second line in Eq.~\eqref{eq:LP_J} makes it clear that the ISP variables are actually not needed to express the integrals. Note that the original Feynman integrals are given by
\begin{equation}
    I(\nu_1,\cdots,\nu_n) = e^{\epsilon\gamma_E L} \, \frac{(-1)^\nu \, \Gamma(d/2)}{\Gamma\left((L+1)d/2-\nu\right)} \left( \prod_{i,\nu_i > 0} \frac{1}{\Gamma(\nu_i)} \right) J(\nu_1,\cdots,\nu_n) \,.
\end{equation}
We want to establish linear relations among these integrals, and reduce them to master integrals of the family. For that we will employ the method of intersection theory and relative cohomology.

\subsection{Intersection theory}

The intersection theory was introduced into Feynman integrals in \cite{Mastrolia:2018uzb}. We consider a family of integrals
\begin{equation}
\int_{\mathcal{C}_R} u \, \varphi_L \,,
\label{eq:hyper}
\end{equation}
where $u$ is a multivalued function called \textit{twist}, which vanishes on the boundary of the integration domain: $u(\partial \mathcal{C}_R) = 0$, and $\varphi_L$ is a differential $n$-form, where $n$ is the number of integration variables. The fact that $u$ vanishes on the boundary means that the integral of a total differential vanishes:
\begin{equation}
	\int_{\mathcal{C}_R} \dif \left( u \, \xi \right) = 0 \,,
\end{equation}
where $\xi$ is an arbitrary $(n-1)$-form. This then leads to the integration-by-parts (IBP) equivalence relations
\begin{equation}
	\varphi_L \sim \varphi_L + \nabla_\w \xi \,,
\end{equation}
where the covariant derivative is defined by
\begin{equation}
\nabla_\w \equiv \dif + \w \wedge \,,
\end{equation}
with the connection $1$-form $\w \equiv \dif \log(u)$.

Instead of working with equivalence classes of $n$-forms, one may also define equivalence classes of integration domains. Given the twist $u$, the central idea of the intersection theory is to treat the equivalence class $|\mathcal{C}_R]$ of integration domains as an element (cycle) in a twisted homology group, and the equivalence classes $\bra{\varphi_L}$ of $n$-forms as an element (cocycle) in a twisted cohomology group $H_\w^n$. The integral in Eq.~\eqref{eq:hyper} can then be interpreted as a non-degenerate bilinear form (see \cite{zbMATH05821564} for more details):
\begin{equation}
	\left \langle \varphi_L \middle | \mathcal{C}_R \right ] \equiv \int_{\mathcal{C}_R} u \, \varphi_L \,.
\end{equation}
With the help of the Poincaré duality, one can introduce dual integrals of the form
\begin{equation}
	\left [ \mathcal{C}_L \middle | \varphi_R \right \rangle \equiv \int_{\mathcal{C}_L} u^{-1} \, \varphi_R \,,
\end{equation}
where $[\mathcal{C}_L|$ is an element of the dual homology group, and $\ket{\varphi_R}$ is an element of the dual cohomology group $H_{-\w}^n$, respectively. In particular, the dual vector $\ket{\varphi_R}$ is the equivalence class of $n$-forms under the equivalence relation
\begin{equation}
\varphi_R \sim \varphi_R + \nabla_{-\w} \xi \,.
\end{equation}

Notably, the dimension of the cohomology group as a vector space corresponds to the number of MIs before considering possible symmetry relations among them. In the framework of Morse theory, this dimension can be determined by counting the critical points of the Morse height function, where $\log(u)$ serves as a canonical example of such a function \cite{Lee:2013hzt}:
\begin{equation}
\dim H_{\pm\omega}^n = \# \{ \text{zeroes of $\omega$} \} \, .
\end{equation}
Here, it is assumed that all possible singularities of the differential forms $\varphi$ are regularized by the twist $u$. If not, one needs to introduce extra regulators when computing the number of critical points. Note that there are freedom in the choice of basis vectors. In practice, we usually choose monomials or $\dif \log$-forms due to their algebraic and analytic simplicity.

Due to the isomorphism between $H_\w^n$ and $H_{-\w}^n$, one can introduce a non-degenerate bilinear pairing between (bra) vectors and dual (ket) vectors, known as the \textit{intersection number} $\braket{\varphi_L | \varphi_R}_\w$, that can be calculated using the methods outlined in \cite{Frellesvig:2019kgj, Frellesvig:2020qot, Brunello:2023rpq, Brunello:2024tqf}. This can be used to perform integral reduction: choosing a basis $\{\bra{e_i}\}$ of the bra vectors and a basis $\{\ket{h_i}\}$ of the ket vectors, any bra vector $\bra{\varphi_L}$ can be expressed as a linear combination:
\begin{equation}
\Bra{\varphi_L} = \sum_{i,j} \Braket{\varphi_L | h_i}_\w \left( \bm{C}^{-1} \right)_{ij} \Bra{e_j} \,,
\end{equation}
where $\bm{C}^{-1}$ is the inverse of the matrix $\bm{C}$ with matrix elements
\begin{equation}
\bm{C}_{ij} \equiv \Braket{e_i | h_j}_{\w} \,.
\end{equation}

In the definition and evaluation of intersection numbers, the singularities of the connection $\omega$ and those of the $n$-forms play an important role. We denote the set of singularities of $\omega$ as $\mathcal{P}_{\w}$ (including infinity). In general, it can happen that some  singularities of the $n$-forms $\varphi_L$ and $\varphi_R$ are not contained in $\mathcal{P}_{\w}$. In that case, one needs to introduce extra regulators into the connection (similar to the counting of dimensions discussed previously), and take the regulators to zero in the end of the calculation. With the regulators in place, the intersection numbers are then given by\footnote{Technically, one needs to reverse the order of the variables in $\varphi_L$ or $\varphi_R$. This may lead to an extra minus sign in the formula.}
\begin{equation}
	\braket{\varphi_L | \varphi_R}_\w \equiv \frac{1}{(2\pi i)^n} \int_{\mathcal{C}} \iota(\varphi_L) \wedge \varphi_R = \frac{(-1)^n}{(2\pi i)^n} \int_{\mathcal{C}} \varphi_L \wedge \iota(\varphi_R) \,,
	\label{eq:int_num}
\end{equation}
where the integration domain is $\mathcal{C}=\mathbb{C}^n \setminus \mathcal{P}_{\w} \cup \mathcal{D}$, with $\mathcal{D}$ being the set of singularities of the $n$-forms, while $\iota(\varphi_L)$ and $\iota(\varphi_R)$ are compactly-supported representatives of $\bra{\varphi_L}$ and $\ket{\varphi_R}$, respectively. Taking the univariate case as an example, these compactly-supported representatives can be obtained using
\begin{equation}
	\iota(\varphi_L) = \varphi_L-\nabla_{\w}(h \psi_L) \,, \quad \iota(\varphi_R) = \varphi_L-\nabla_{-\w}(h \psi_R) \,,
\end{equation}
where $h$ is a combination of Heaviside functions given by
\begin{equation}
	h = \sum_{p \in \mathcal{P}_{\w}}(1-\theta_{x,p}) \,, \quad \theta_{x,p} = \theta(|x-p|-\epsilon) \,,
	\label{eq:theta}
\end{equation}
and the functions $\psi_L$ and $\psi_R$ are the solutions to the following differential equations:
\begin{equation}
	\nabla_{\w} \psi_L = \varphi_L \,, \quad \nabla_{-\w} \psi_R = \varphi_R \, .
\end{equation}
Note that if $\varphi_L$ or $\varphi_R$ contains singularities which are not regularized by $\omega$, the above equations may have no solution. That's the reason why the extra regulators are necessary. It can be seen that the compactification guarantees that the integrals in Eq.~\eqref{eq:int_num} are well-defined since the neighborhoods of all singularities are avoided. Therefore, the intersection number should actually be regarded as a bilinear map $\Braket{\bullet | \bullet}: H_{\w}^n \times H_{-\w,c}^n \to \mathbb{C}$ or $\Braket{\bullet | \bullet}: H_{\w,c}^n \times H_{-\w}^n \to \mathbb{C}$, where the subscript $c$ denotes compactification.

\subsection{Relative cohomology in the Baikov representation}

In the existing approaches of using intersection theory for Feynman integral reduction, one usually works in the Baikov representation. In the Baikov representation, the integration variables are the propagator denominators: $z_i = D_i$, and the $n$-forms $\varphi_L$ corresponding to Feynman integrals contain factors like $1/z_i^{\nu_i}$. It is often the case that the singularities at $z_i = 0$ are not regularized by the connection $\omega$, and one needs to introduce extra regulators as discussed previously. In practice, this increases the complexity of the calculation.

In \cite{Matsumoto:2018igm}, it is observed that the cohomology group of compactly-supported forms is isomorphic to a so-called \textit{relative} cohomology group. In \cite{Caron-Huot:2021xqj, Caron-Huot:2021iev}, it is then proposed that one can replace the dual cohomology group $H_{-\w,c}^n$ by the corresponding relative cohomology group $H^n_{-\w}(\mathbb{C}^n \setminus \mathcal{P}_\omega, \mathcal{D})$. We recall that $\mathcal{D}$ is the set of singularities in the $n$-forms, which are called relative boundaries in the language of relative cohomology. On the contrary, the set $\mathcal{P}_w$ contains the so-called twisted boundaries. For the Baikov representation, the set of relative boundaries is just $\mathcal{D} = \bigcup_{i} \{z_i = 0\}$. The concept of relative cohomology allows for the computation of intersection numbers without introducing extra regulators, hence simplifies the problem of Feynman integral reduction.

For the purpose of integral reduction, the most important concept in relative cohomology is the boundary-supported form, or $\delta$-form for short. These forms live on the relative boundaries. They give rise to equivalent contributions to the intersection numbers as introducing extra regulators for these boundaries \cite{Brunello:2023rpq}. Without loss of generality, we consider a $\delta$-form living on the boundary $z_{m+1}=\cdots=z_n = 0$. It is defined by
\begin{equation}
	\label{eq:leray}
	\delta_{z_{m+1},\cdots,z_n} \phi_R = \frac{u(z_1,\cdots,z_m,z_{m+1},\cdots,z_n)}{u(z_1,\cdots,z_m,0,\cdots,0)} \, \phi_R \wedge \dif \theta_{z_{m+1},0} \wedge \cdots \wedge \dif \theta_{z_{n},0} \,,
\end{equation}
where the $\theta$-function is defined in Eq.~\eqref{eq:theta}, and $\phi$ is an $m$-form of the variables $z_1,\cdots,z_m$. Here, the $\delta$ operator is a realization of the Leray coboundary map (see, e.g., \cite{pham2011singularities} for a comprehensive review). The differential form $\dif \theta_{z,0}$ should be regarded as a distribution. When acting on a test function $f$, it gives
\begin{equation}
	\dif\theta_{z,0} \left[ f \right] = - \lim\limits_{\epsilon\to 0}\int_{-\infty}^{+\infty} f'(z) \, \theta(z - \epsilon) \, \dif z \,.
	\label{eq:dtheta}
\end{equation}

 The intersection number between a bra vector $\bra{\varphi_L}$ and a $\delta$-form is given by
\begin{align}
	\Braket{\varphi_L | \delta_{z_{m+1},\cdots,z_n}\phi_R}_\w  &= \frac{(-1)^n}{(2\pi i)^n} \int_{\mathcal{C}} \varphi_L \wedge \iota (\delta_{z_{m+1},\cdots,z_n} \phi_R) \nonumber
	\\
	&= \Braket{\Res_{z_{m+1} = \cdots = z_n = 0} \frac{u}{u_0} \varphi_L | \phi_R }_{\w_0} \,,
	\label{eq:relative_intersection}
\end{align}
where the $\iota$-map only deals with the variables $z_{1}, \cdots, z_{m}$, and
\begin{equation}
	u_0 = u(z_1,\cdots,z_m,0,\cdots,0) \,, \quad \w_0 = \dif \log(u_0) \,. 
\end{equation}

Note that if $\varphi_L$ has only simple poles in the variables $z_{m+1},\cdots,z_n$, the multivariate residue is trivial to take, and the factor $u/u_0$ has no effect in the residue. However, since $\varphi_L$ represents the Feynman integrals to reduce, it generally contains higher-order poles corresponding to, e.g., doubled propagators. In this case, one may convert $\varphi_L$ to another representative with only simple poles in the same equivalence class, by performing IBPs on the variables $z_{m+1},\cdots,z_n$.

\subsection{Relative cohomology in the Feynman parametrization}

We now introduce the main novel idea of this work. Looking at the Feynman integrals in the LP representation \eqref{eq:LP_J}, we see that the integrands automatically have the structure of a relative cohomology. The $n$-forms that are not boundary-supported correspond to Feynman integrals in the top sector, while those living on the boundary correspond to integrals in sub-sectors. Therefore, instead of introducing $\delta$-forms in the dual cohomology, we will have $\delta$-forms in the bra vectors $\bra{\varphi_L}$, which represents equivalence classes of Feynman integrals and belongs to the relative cohomology group $H_\w^n(\mathbb{C}^n \setminus \mathcal{P}_\w, \mathcal{D})$. Here, the set of relative boundaries is just the zero locus of Feynman parameters: $\mathcal{D} = \bigcup_i \{ x_i = 0 \}$.

In our approach, the dual cohomology is now the usual cohomology group $H_{-\w}^n$. The $n$-forms in the dual cohomology will contain factors such as $1/x_i$, that are singular at the relative boundaries. These singular terms are necessary to correctly pick up the contributions from the relative boundaries to the intersection numbers.

At this point, it is worth discussing again the counting of dimensions in the case of relative cohomology. Since the relative cohomology groups are isomorphic to the usual cohomology groups, the dimensions can be computed as before by introducing regulators for the relative boundaries. Alternatively, one may perform the counting by combining the number of MIs living in the bulk (top-sector) and those living on each relative boundary (sub-sectors). In the latter approach, one needs to be careful with the possible ``magic-relations'' that relates integrals in different sub-sectors.

Without loss of generality, an $n$-form in the bra vector has the form
\begin{equation}
	\delta_{x_{m+1},\cdots,x_n} \phi_L = \frac{u(x_1,\cdots,x_m,0,\cdots,0)}{u(x_1,\cdots,x_m,x_{m+1},\cdots,x_n)} \, \phi_L \wedge \dif \theta_{x_{m+1},0} \wedge \cdots \wedge \dif \theta_{x_{n},0} \,.
	\label{eq:delta_form}
\end{equation}
The Riesz–Markov–Kakutani representation theorem guarantees the uniqueness of the measures in the $\dif\theta$ form (see \cite{rudin1991functional} for more mathematical details). Hence, the action of $\dif\theta$ on a test function $f$, Eq.~\eqref{eq:dtheta}, is equivalent to the integral of $f$ with a $\delta$-distribution inserted as a Radon measure as in Eq.~\eqref{eq:delta_form}. This exactly corresponds to what we have in the Feynman parametrization. For convenience, we will write the above $\delta$-form as
\begin{equation}
	\delta_{x_{m+1},\cdots,x_n} \phi_L = \phi_L \wedge \delta(x_{m+1}) \dif x_{m+1} \wedge \cdots \wedge \delta(x_{n}) \dif x_{n} \,.
\end{equation}
For the reduction of these integrals, we choose a basis $\{\ket{h_i}\}$ in the dual cohomology, and evaluate the intersection numbers
\begin{equation}
    \Braket{\delta_{x_{m+1},\cdots,x_n} \phi_L | h_i}_\w = \Braket{\phi_L | \Res_{x_{m+1} = \cdots = x_n = 0} \frac{u_0}{u} h_i}_{\w_0} \,.
\end{equation}
Note that we can always choose the dual basis $\{\ket{h_i}\}$ to have at most simple poles at the relative boundaries, and therefore the above intersection numbers can be simplified as:
\begin{equation}
    \Braket{\delta_{x_{m+1},\cdots,x_n} \phi_L | h_i}_\w = \Braket{\phi_L |\Res_{x_{m+1} = \cdots = x_n = 0} h_i}_{\w_0} \,.
\end{equation}
This can be contrasted with the other approaches with $\delta$-forms in the dual basis, where the integrand $\varphi_L$ for reduction can have higher-order poles at the relative boundaries. 

After taking the residues, we still need to compute the ``normal'' intersection numbers between $\phi_L$ and the residues of $h_i$. For completeness and for later applications, we briefly outline the relevant procedure in the following. For more details, we refer the readers to Refs.~\cite{Frellesvig:2019kgj, Frellesvig:2020qot, Brunello:2023rpq, Brunello:2024tqf}.

The computation of the multivariate intersection numbers usually proceeds variable-by-variable. For that we first choose an order of the variables $\bm{z}=\{z_1,\cdots,z_N\}$, where we assume that there are $N$ variables involved in the intersection number. Furthermore, we use the boldface letter $\bm{n}$ to denote the sequence $\{1,\cdots,n\}$, where $1 \leq n \leq N$. The bra vector $\bra{\varphi_L^{(\bm{n})}}$ at layer $\bm{n}$ is the equivalence class of $n$-forms
\begin{equation}
\varphi_L^{(\bm{n})} = \hat{\varphi}_L^{(\bm{n})}(\bm{z}) \, \dif z_1 \wedge \dif z_2 \wedge \cdots \wedge \dif z_n \, ,
\end{equation}
with the connection given by
\begin{equation}
\w^{(\bm{n})} = \hat{\w}_1(\bm{z}) \, \dif z_1 + \hat{\w}_2(\bm{z}) \, \dif z_2 + \cdots + \hat{\w}_n(\bm{z}) \, \dif z_n \, , \quad \hat{\omega}_i(\bm{z}) = \frac{\partial}{\partial z_i} \log(u) \,.
\end{equation}
Note that the variables $z_{n+1},\cdots,z_{N}$ are treated as constants here. The ket vector $\ket{\varphi_R^{(\bm{n})}}$ is defined similarly, with the connection $-\w^{(\bm{n})}$. The intersection number between $\bra{\varphi_L^{(\bm{n})}}$ and $\ket{\varphi_R^{(\bm{n})}}$ can then be computed recursively as follows. We choose a basis $\bra{e_i^{(\bm{n-1})}}$ and a dual basis $\ket{h_i^{(\bm{n-1})}}$ for the layer $\bm{n-1}$. The metric matrix for the layer $\bm{n-1}$ is given by
\begin{equation}
	\left( \bm{C}_{(\bm{n-1})} \right)_{ij} \equiv \Braket{e_i^{(\bm{n-1})} | h_j^{(\bm{n-1})}}_{\w^{(\bm{n-1})}} \,.
\end{equation}
These ingredients can be used to compute the coefficients
\begin{equation}
	\bra{\varphi_{L,j}^{(n)}} = \braket{\varphi_L^{(\bm{n})} | h_i^{(\bm{n-1})}}_{\w^{(\bm{n-1})}} \big( \bm{C}_{(\bm{n-1})}^{-1} \big)_{ij} \,, \quad \ket{\varphi_{R,j}^{(n)}} = \big( \bm{C}_{(\bm{n-1})}^{-1} \big)_{ji} \braket{e^{(\bm{n-1})}_i | \varphi_R^{(\bm{n})}}_{\w^{(\bm{n-1})}} \, ,
\end{equation}
where repeated indices are summed over. The bra and ket notation means that these coefficients are equivalence classes of vector-valued one-forms in the variable $z_n$. For the dual coefficients, the relevant connection matrix is given by
\begin{equation}
    \Omega_{ij}^{\vee(n)} = \big( \bm{C}_{(\bm{n-1})}^{-1} \big)_{ik} \Braket{e_k^{(\bm{n-1})} | (\partial_{z_n}-\hat{\w}_n) h_j^{(\bm{n-1})}}_{\w^{(\bm{n-1})}} \, .
\end{equation}
The intersection number at layer $\bm{n}$ is finally given by
\begin{align}
    \Braket{\varphi_L^{(\bm{n})} | \varphi_R^{(\bm{n})}}_{\w^{(\bm{n})}} &= -\sum_{p \in \mathcal{P}_n} \Res_{z_n = p} \big( \varphi_{L,i}^{(n)} \left( \bm{C}_{(\bm{n-1})} \right)_{ij} \psi_{R,j}^{(n)} \big) \, ,
\end{align}
where $\mathcal{P}_n$ is the set of singularities of $\Omega_{ij}^{\vee(n)}$, and $\psi_{R,j}^{(n)}$ is the solution to the equation
\begin{equation}
    \partial_{z_n} \psi_{R,j}^{(n)} + \Omega^{\vee(n)}_{ji} \psi^{(n)}_{R,i} = \hat{\varphi}_{R,j}^{(n)} \,.
\end{equation}

\section{Examples}\label{sec:Examples}

\subsection{One-loop bubble}

\begin{figure}[!htbp]
	\centering
	\includegraphics[width=0.4\linewidth]{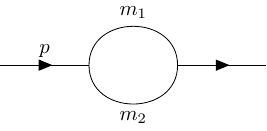}
	\caption{The one-loop bubble diagram with two different masses $m_1$ and $m_2$.}
	\label{fig:bubble2mass}
\end{figure}

As a simple example, we consider the one-loop bubble diagram shown in Figure \ref{fig:bubble2mass}. The kinematic variables are $m_1^2$, $m_2^2$ and $p^2=s$.
The integrals in this family can be represented by
\begin{equation}
	J_{\text{Bub}} (\nu_1, \nu_2) = \int_0^\infty \left( \prod_{i} x_i^{\nu_i-1} \, \dif x_i \right) G^{-d/2} \, ,
\end{equation}
where the LP polynomial is
\begin{equation}
	G = \mathcal{U} + \mathcal{W} = x_1 + x_2 + m_1^2 x_1^2 + m_2^2 x_2^2 + m_1^2 x_1 x_2 + m_2^2 x_1 x_2 - 
	s x_1 x_2 \, .
\end{equation}

To proceed, we choose the variable order as $\{x_1,x_2\}$, i.e., $\bm{1} \equiv \{1\}$ is the inner layer, while $\bm{2} \equiv \{1,2\}$ is the outermost layer. As discussed in the previous section, there are two equivalent methods for counting the dimension for each layer, with or without extra regulators. For the inner layer, the critical condition with the regulator is given by
\begin{equation}
	\omega^{(\bm{1})}_{\rho} = \frac{\partial}{\partial x_1} \log \left( x_1^{\rho_1}  G^{-d/2} \right) \dif x_1 = 0 \, ,
\end{equation}
where the subscript $\rho$ signals that the connection form is regularized. The above equation has two solutions, meaning that the dimension of the inner layer is $\nu^{(\bm{1})} = 2$. Alternatively, we can perform the counting in each sectors separately, and sum the results. For the inner layer, there are two sectors: the top-sector $1$ and the sub-sector $\delta(x_1)$. The dimension for the sub-sector $\delta(x_1)$ is simply $1$, since the only integration variable $x_1$ is fixed by the $\delta$ function. For the top-sector, the critical condition reads
\begin{equation}
	\omega^{(\bm{1})} = \frac{\partial}{\partial x_1} \log \left( G^{-d/2} \right) \dif x_1 = 0 \,,
\end{equation}
which has one solution. Therefore, the total dimension is $\nu^{(\bm{1})} = 2$, in agreement with the result with regulators. For the outer layer, the counting proceeds similarly. The regularized critical condition reads
\begin{equation}
	\omega^{(\bm{2})} = \left( \dif x_1 \, \frac{\partial}{\partial x_1} + \dif x_2 \, \frac{\partial}{\partial x_1} \right) \left( \log x_1^{\rho_1} x_2^{\rho_2} G^{-d/2} \right) = 0 \,,
\end{equation}
which has 3 solutions. In the sector-by-sector counting scheme, we need to consider 3 non-zero sectors $1$, $\delta(x_1)$ and $\delta(x_2)$. The number of MIs in each sector is again $1$ and the total dimension is $\nu^{(\bm{2})} = 3$.

The basis of the inner layer should be chosen according to the dimensions of each sector. For example, we may choose
\begin{equation}
	\hat{e}_1^{(\bm{1})} = 1 \,, \quad \hat{e}_2^{(\bm{1})} = \delta(x_1) \,.
\end{equation}
For the outer layer, we can choose the basis as
\begin{equation}
	\hat{e}_1 = \hat{e}_1^{(\bm{2})} = 1 \,, \quad \hat{e}_2 = \hat{e}_2^{(\bm{2})} = \delta(x_1) \,, \quad \hat{e}_3 = \hat{e}_3^{(\bm{2})} = \delta(x_2) \,.
\end{equation}
They correspond to the MIs $J_{\text{Bub}}(1,1)$, $J_{\text{Bub}}(0,1)$ and $J_{\text{Bub}}(1,0)$. The dual bases can be chosen correspondingly with the following rule: for each $\delta(x_i)$, the dual vector needs to contain a factor of $1/x_i$ in order to correctly account for the relative boundaries. For example, the dual bases for the inner and outer layers can be chosen as
\begin{equation}
	\hat{h}^{(\bm{1})}_i \in \left\lbrace 1, \frac{1}{x_1} \right\rbrace , \quad \hat{h}_i = \hat{h}^{(\bm{2})}_i \in \left\lbrace 1, \frac{1}{x_1}, \frac{1}{x_2} \right\rbrace \, .
\end{equation}

To demonstrate our method outlined in the previous section, let us decompose the target integral $I_{\text{Bub}} (1,2)$ into the 3 MIs:
\begin{equation}
	I_{\text{Bub}} (1,2) = c_1 \, I_{\text{Bub}} (1,1) + c_2 \, I_{\text{Bub}} (0,1) + c_3 \, I_{\text{Bub}} (1,0) \,.
\end{equation}
To do that we first need to compute the metric matrices $\bm{C}_{(\bm{n})}$. For the inner layer, we find
\begin{equation}
	\bm{C}_{(\bm{1})} =
	\begin{pmatrix}
		-\frac{d \left[ x_2^2 \lambda(m_1^2,m_2^2,s) -2(m_1^2-m_2^2+s)x_2 +1 \right]}{4 (d-1) (d+1) m_1^4} & \frac{(m_1^2+m_2^2-s)x_2 +1}{2 (d-1) m_1^2} \\
		0 & 1
	\end{pmatrix} \, ,
\end{equation}
where we have introduced the Källén function $\lambda(x,y,z)$ for convenience:
\begin{equation}
	\lambda(x,y,z) = x^2+y^2+z^2-2xy-2yz-2zx \, .
\end{equation}
The connection matrix for the second layer can then be computed as
\begin{equation}
	\Omega^{\vee(2)} =
	\begin{pmatrix}
		\frac{(d+1)\left[\lambda(m_1^2,m_2^2,s)x_2-m_1^2+m_2^2-s\right]}{\lambda(m_1^2,m_2^2,s)x_2^2-2(m_1^2-m_2^2+s)x_2+1} & -\frac{(d+1)\left[(m_1^2- m_2^2- s)x_2-1\right]m_1^2}{(x_2 + m_2^2 x_2^2) \left[\lambda(m_1^2,m_2^2,s)x_2^2-2(m_1^2-m_2^2+s)x_2+1\right]} \\
		0 & \frac{d (1 + 2 m_2^2 x_2)}{2 x_2 (1 + m_2^2 x_2)}
	\end{pmatrix} \, .
\end{equation}
We can then compute the metric matrix for the second layer, whose non-zero entries are
\begin{eqnarray}
	\left( \bm{C}_{(\bm{2})} \right)_{1,1} &=& \dfrac{4 s^2}{(4-d^2) \lambda(m_1^2,m_2^2,s)^3} \,, \nonumber
	\\
	\left( \bm{C}_{(\bm{2})} \right)_{1,2} &=& \dfrac{(s-m_1^2 + m_2^2)[d(\lambda(m_1^2,m_2^2,s)-4m_2^2 s)+4m_2^2 s]}{4 (d-2) (d^2-1) m_2^4 \lambda(m_1^2,m_2^2,s)^2} \,, \nonumber
	\\
	\left( \bm{C}_{(\bm{2})} \right)_{1,3} &=& \dfrac{(s+m_1^2 - m_2^2)[d(\lambda(m_1^2,m_2^2,s)-4m_1^2 s)+4m_1^2 s]}{4 (d-2) (d^2-1) m_1^4 \lambda(m_1^2,m_2^2,s)^2} \,, \nonumber
	\\
	\left( \bm{C}_{(\bm{2})} \right)_{2,2} &=& -\dfrac{d}{4 (d^2-1) m_2^4} \,, \nonumber
	\\
	\left( \bm{C}_{(\bm{2})} \right)_{3,3} &=& -\dfrac{d}{4 (d^2-1) m_1^4} \,.
\end{eqnarray}

Our target integral corresponds to the two-form $\varphi_L = x_2 \, \dif x_1 \wedge \dif x_2$. Its intersection numbers with the dual basis are given by
\begin{equation}
	\Braket{\varphi_L | h^{(\bm{2})}_i} =
	\begin{pmatrix}
		\frac{m_1^2-m_2^2+s}{\lambda(m_1^2,m_2^2,s)}, & -\frac{m_1^2+m_2^2-s}{2(d-3)m_2^2 \lambda(m_1^2,m_2^2,s)}, & \frac{1}{(d-3) \lambda(m_1^2,m_2^2,s)}
	\end{pmatrix} \, .
\end{equation}
We multiply the above vector by the matrix $\bm{C}_{(\bm{2})}^{-1}$, and take care of the conversion factors between $I_{\text{Bub}}(\nu_1,\nu_2)$ and $J_{\text{Bub}}(\nu_1,\nu_2)$, and finally arrive at the reduction coefficients
\begin{equation}
	c_1 = -\frac{(d-3)(m_1^2-m_2^2+s)}{\lambda(m_1^2,m_2^2,s)}, \quad c_2=-\frac{(d-2)(m_1^2+m_2^2-s)}{2 m_2^2 \lambda(m_1^2,m_2^2,s)} , \quad c_3 = \frac{d-2}{\lambda(m_1^2,m_2^2,s)} \, .
\end{equation}
They agree with the results of \texttt{Kira}.

We can reduce integrals with numerators as well, according to Eq.~\eqref{eq:LP_J}. Taking the target integral $I_{\text{Bub}}(-1,2)$ as an example, we can write the corresponding two-form as
\begin{equation}
	\varphi_{L}= \frac{d}{2} \frac{x_{2}\left(1+m_{1}^{2}x_{2}+m_{2}^{2}x_{2}-sx_{2}\right)}{G} \, \delta(x_{1})\dif x_{1}\wedge\dif x_{2}\, .
\end{equation}
As mentioned below Eq.~\eqref{eq:ibp_for_num}, we have the polynomial $G$ in the denominator here. Since $G = 0$ is a twisted boundary that is regularized by $u$, this denominator poses no difficulty in the computation of intersection numbers. Repeating the procedure as above, we find
\begin{equation}
	I_{\text{Bub}}(-1,2)=\frac{2m_2^2-\left(d-2\right)\left(m_{1}^{2}-m_{2}^{2}-s\right)}{2m_{2}^{2}} \, I_{\text{Bub}}(0,1) \, ,
\end{equation}
which is again verified by \texttt{Kira}.

\subsection{Two-loop sunrise}

\begin{figure}[!htbp]
	\centering
	\begin{subfigure}{0.4\linewidth}
		\centering
		\includegraphics[width=\linewidth]{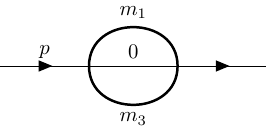}
		\caption{Sunrise sub-topology diagram.}
	\end{subfigure}
	\begin{subfigure}{0.4\linewidth}
		\centering
		\includegraphics[width=\linewidth]{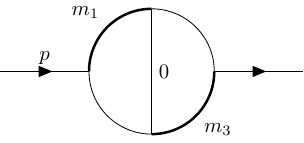}
		\caption{Full two-loop bubble diagram.}
	\end{subfigure}
	\caption{The sunrise integrals with two different internal masses, $m_1$ and $m_3$. The thick lines denote massive propagators, while the thin internal line denotes a massless propagator.}
	\label{fig:sunrise2mass}
\end{figure}

We now consider a two-loop example, the sunrise family regarded as a sub-topology of the two-loop bubble diagram shown in Figure \ref{fig:sunrise2mass}. The propagator denominators for the bubble diagram are given by
\begin{equation}
	D_1=k_1^2-m_1^2 \,, \quad D_2=(k_1-k_2)^2 \,, \quad D_3=(k_2+p)^2-m_3^2 \,, \quad D_4=k_2^2 \,, \quad D_5=(k_1+p)^2 \,.
\end{equation}
The kinematics variables are $m_1^2$, $m_3^2$ and $p^2=s$. The LP polynomial for the full bubble family is
\begin{align}
	G &= (x_{1} x_{2} + x_{1} x_{3} + x_{2} x_{3} + x_{1} x_{4} + x_{2} x_{4} +  x_{2} x_{5} + x_{3} x_{5} + x_{4}  x_{5}) (1+m_1^2 x_1 + m_3^2 x_3) \nonumber
	\\
	&- s \, (x_{1} x_{2} x_{3} + x_{1} x_{3} x_{4} + x_{2} x_{3} x_{4} + x_{1} x_{2} x_{5} +  x_{1} x_{3} x_{5} + x_{1} x_{4} x_{5} + x_{2} x_{4} x_{5} + x_{3} x_{4} x_{5}) \, .
\end{align}
Since we are interested in the sunrise sub-topology, we only need to compute intersection numbers with the LP polynomial restricted to the relative boundary $x_4 = x_5 = 0$:
\begin{equation}
	G_0 = x_1 x_2 + x_1 x_3 + x_2 x_3 + m_1^2 x_1^2 (x_2 + x_3) + (m_1^2 + m_3^2 -s) x_1 x_2 x_3 + m_3^2 x_3^2(x_1+x_2) \,.
\end{equation}
In the context of the sunrise sub-topology, $D_4$ and $D_5$ are ISPs. When performing the reduction in the momentum representation or the Baikov representation, ISPs are generically required for two-loop integrals and beyond. On the contrary, in the Feynman parametrization, ISPs do not need to appear in the computation of intersection numbers, as discussed around Eqs.~\eqref{eq:LP_J} and \eqref{eq:LP_G0}. This leads to simplification of the reduction procedure.

The counting of the dimensions for each layer proceeds similarly as the one-loop case. We choose the variable order as $\{x_1,x_2,x_3\}$. For the first layer, there are again two sectors: the top-sector $1$ and the sub-sector $\delta(x_1)$, each with dimension $1$. For the second layer, there are two irreducible sectors: the top-sector $1$ and the sub-sector $\delta(x_2)$. Note that the sub-sector $\delta(x_1)$ in the second layer is reducible, since after taking $x_1 \to 0$, $G_0$ becomes a linear polynomial of $x_2$, which leaves no irreducible integrals. This is related to the fact that the second propagator is massless. The critical condition for the sub-sector $\delta(x_2)$ is given by
\begin{equation}
	\omega^{(\bm{2})}_{\delta(x_2)} = \dif x_1 \, \frac{\partial}{\partial x_1} \left( \log G^{-d/2} \right)_{x_2 = 0} = 0 \,,
\end{equation}
which has one solution. For the top-sector, the critical condition is given by
\begin{equation}
	\omega^{(\bm{2})} = \left( \dif x_1 \, \frac{\partial}{\partial x_1} + \dif x_2 \, \frac{\partial}{\partial x_1} \right) \left( \log G^{-d/2} \right) = 0 \,,
\end{equation}
which has two solutions. Therefore, the dimension for the second layer is $\nu^{(\bm{2})} = 3$, which coincides with the result from the regularized twist. Finally, for the outer layer we have $\nu^{(\bm{3})} = 4$. The bases for all layers are chosen as the following:
\begin{equation}\label{eq:sun_bases}
	\hat{e}^{(\bm{1})}_i \in \left\lbrace 1, \delta(x_1) \right\rbrace , \quad \hat{e}^{(\bm{2})}_i \in \left\lbrace 1, x_2, \delta(x_2) \right\rbrace \,, \quad \hat{e}_i = \hat{e}^{(\bm{3})}_i \in \left\lbrace 1, x_1, x_3, \delta(x_2) \right\rbrace .
\end{equation}
The basis for the outer layer corresponds to the MIs $I_{\mathrm{Sun}} (1,1,1,0,0)$, $I_{\mathrm{Sun}} (2,1,1,0,0)$, $I_{\mathrm{Sun}} (1,1,2,0,0)$ and $I_{\mathrm{Sun}} (1,0,1,0,0)$. The corresponding dual bases are given by
\begin{equation}
	\hat{h}^{(\bm{1})}_i \in \left\lbrace 1, \frac{1}{x_1} \right\rbrace , \quad \hat{h}^{(\bm{2})}_i \in \left\lbrace 1, x_2, \frac{1}{x_2} \right\rbrace , \quad \hat{h}_i = h^{(\bm{3})}_i \in \left\lbrace 1, x_1,  x_3, \frac{1}{x_2} \right\rbrace .
\end{equation}

We begin by considering the reduction of integrals without numerators, i.e., $a_4 = a_5 = 0$. As an example, we reduce the target integral $I_{\mathrm{Sun}} (1,2,1,0,0)$ into the 4 MIs:
\begin{align}
	I_{\mathrm{Sun}} (1,2,1,0,0) &= c_1 I_{\mathrm{Sun}} (1,1,1,0,0) + c_2 I_{\mathrm{Sun}} (2,1,1,0,0) \nonumber
	\\
	&+ c_3 I_{\mathrm{Sun}} (1,1,2,0,0) + c_4 I_{\mathrm{Sun}} (1,0,1,0,0) \, .
\end{align}
Repeating the procedures as in the one-loop example, we find
\begin{eqnarray}
	c_1 &=& \dfrac{(d-3)(3d-8)(m_1^2+m_3^2-s)}{(d-4)\lambda(m_1^2,m_3^2,s)} \,, \nonumber
	\\
	c_2 &=& -\dfrac{4(d-3)m_1^2(m_1^2-s)}{(d-4)\lambda(m_1^2,m_3^2,s)} \,, \nonumber
	\\
	c_3 &=& -\dfrac{4(d-3)m_3^2(m_3^2-s)}{(d-4)\lambda(m_1^2,m_3^2,s)} \,, \nonumber
	\\
	c_4 &=& -\dfrac{(d-2)^2}{(d-4)\lambda(m_1^2,m_3^2,s)} \,.
\end{eqnarray}
They again agree with the results of \texttt{Kira}.

For integrals featuring numerators, we examine a representative case involving the decomposition of $I_{\mathrm{Sun}} (1,1,1,-1,-1)$. Following Eqs.~\eqref{eq:LP_J} and \eqref{eq:LP_G0}, the corresponding differential form is expressed as
\begin{align}
	\varphi_{L} &= \left[ G^{d/2} \frac{\partial^2}{\partial x_4 \partial x_5}  G^{-d/2} \right]_{x_4=x_5=0} \, \dif x_{1}\wedge\dif x_{2} \wedge \dif x_3 \nonumber
	\\
	&= \frac{d}{4 G_0^2} \Big[ d (x_{1} + x_{2}) (x_{2} + x_{3}) (1 + m_{1}^{2} x_{1} - s x_{1} + m_{3}^{2} x_{3}) (1 + m_{1}^{2} x_{1} - s x_{3} + m_{3}^{2} x_{3}) \nonumber
	\\
	&\hspace{6em} + 2 (s x_{1} x_{3} + x_{2} (1 + m_{1}^{2} x_{1} + m_{3}^{2} x_{3}))^2\Big]  \, \dif x_{1}\wedge\dif x_{2} \wedge \dif x_3 \, .
	\label{eq:phiL_numerator}
\end{align}
The basis and dual basis for each layer remain unchanged. By calculating intersection numbers it reveals that
\begin{align}
	I_{\mathrm{Sun}} (1,1,1,-1,-1) &= c'_1 I_{\mathrm{Sun}} (1,1,1,0,0) + c'_2 I_{\mathrm{Sun}} (2,1,1,0,0) \nonumber
	\\
	&+ c'_3 I_{\mathrm{Sun}} (1,1,2,0,0) + c'_4 I_{\mathrm{Sun}} (1,0,1,0,0) \, .
\end{align}
with
\begin{eqnarray}
	c'_1 &=& \dfrac{(6-2d)(m_1^4+m_3^4)+(17d-36)m_1^2 m_3^2 + (16-7d)(m_1^2 s + m_3^2 s) + (d-2)s^2}{9d-12} \,, \nonumber
	\\
	c'_2 &=& \dfrac{2m_1^2(m_1^2-s)(m_1^2-5 m_3^2 +s)}{9d-12} \,, \nonumber
	\\
	c'_3 &=& \dfrac{2m_3^2(m_3^2-s)(m_3^2-5 m_1^2 +s)}{9d-12} \,, \nonumber
	\\
	c'_4 &=& \dfrac{(d-2)(m_1^2+m_3^2)+(10d-14)s}{9d-12} \,.
\end{eqnarray}
Note that one may choose integrals with numerators as master integrals as well, by selecting the form \eqref{eq:phiL_numerator} into the basis.

We have repeatedly emphasized the advantages of using the LP parametrization over the Baikov representations, and it is a good time to show this explicitly. The first advantage is the reduced number of variables which leads to reduced number of layers in the computation of intersection numbers. In the setup phase of the computation, one needs to calculation $\nu^{(\bm{i})} \times \nu^{(\bm{i})}$ $i$-variables intersection numbers at layer $\bm{i}$ to determine the matrix $\bm{C}_{(\bm{i})}$ and its inverse, with additional calculations for the connection matrix $\Omega^{\vee (i)}$. In the reduction phase, one needs to compute the projections of the target integrals onto the dual basis vectors, which again requires the calculation of $n$-variables intersection numbers where $n$ is the total number of layers. Therefore, the reduced number of variables is already a big computational advantage. Using the two-loop sunrise family as an example, there are only three layers in the LP parameterization, while all five layers need to be considered in the Baikov representation\footnote{One may employ the so-called loop-by-loop Baikov representation with four variables. However, that would leads to new problems such as increased dimensionalities due to the presence of non-Feynman-integrals in the integral family. This has been thoroughly discussed in \cite{Chen:2022lzr}.}.

The second advantage of the LP parameterization is the simplicity of the LP polynomials. For comparison, the Baikov polynomial for the two-loop sunrise family is given by
\begin{align}
	\mathcal{B}&= m_1^2 m_3^2 \left( -2 x_1+x_2-2 x_3+x_4+x_5 - m_1^2 - m_3^2 \right) \nonumber \\
	&+ m_1^2 \left( -x_3^2-2 x_1 x_3+x_2 x_3+x_4 x_3+x_5
	x_3-x_2 x_4+x_4 x_5 - m_1^2 x_3 \right) \nonumber \\
	&+ m_3^2 \left( -x_1^2+x_2 x_1-2 x_3 x_1+x_4 x_1+x_5
	x_1-x_2 x_5+x_4 x_5 - m_3^2 x_1 \right) \nonumber \\
	&+ s \left[ m_1^2 m_3^2 + m_1^2 (x_2 + x_3 - x_5) + m_3^2 (x_1 + x_2 - x_4) - s x_2\right]  \nonumber \\
	&+ s \left(x_1 x_2 -x_2^2 +x_3 x_2+x_4 x_2+x_5x_2+x_1 x_3-x_3 x_4-x_1 x_5+x_4 x_5 \right) \nonumber \\
	&- x_1 x_3 \left( x_1-x_2+x_3-x_4-x_5 \right) + x_4 x_5 \left( x_1+x_2+x_3-x_4-x_5 \right) -x_2 \left( x_1x_4+x_3x_5 \right)   \, .
\end{align}
Due to the fact that there are a lot of manipulation of the polynomials involved in the computation of intersection numbers, the simplicity of the LP polynomial provides an additional advantage over the Baikov one. Finally, one may also compare the dimensions of each layer, since it also affects the number of intersection numbers one needs to calculate. As present earlier, the dimensions in the LP parametrization are $\nu^{(\bm{1})} = 2$, $\nu^{(\bm{2})} = 3$ and $\nu^{(\bm{3})} = 4$, while the dimensions in the Baikov representation are $\nu^{(\bm{1})}=1$, $\nu^{(\bm{2})}=4$, $\nu^{(\bm{3})}=4$, $\nu^{(\bm{4})}=4$ and $\nu^{(\bm{5})}=4$ with the variable order $\{x_5,x_4,x_2,x_3,x_1\}$. In principle, one may also compare the intermediate expressions such as the $\bm{C}_{(\bm{i})}$ and $\Omega^{\vee (i)}$ matrices. These expressions are too lengthy and we do not give them here explicitly.

\subsection{The degenerate limits}
\label{sec:degenerate}

The method of relative cohomology has subtleties in certain degenerate limits \cite{Caron-Huot:2021xqj}. For example, in the one-loop bubble family, if the external momentum becomes light-like, i.e., $p^2 = 0$, the LP polynomial becomes factorized:
\begin{equation}
    G = \left( x_1 + x_2 \right) \left( 1 + x_1 m_1^2 + x_2 m_2^2 \right) .
\end{equation}
In this case, the dimension of the cohomology group decreases by one, since the top-sector now become reducible. Hence, there are only two master integrals instead of three. We have checked that by simply choosing a smaller basis, such as $I_{\text{Bub}}(1,0)$ and $I_{\text{Bub}}(0,1)$, we can correctly perform the reduction as usual. In particular, we can find the correct reduction rule of $I_{\text{Bub}}(1,1)$, which was a master when $p^2 \neq 0$. 

An even trickier degenerate limit is when $m_1 = m_2 = m$, in addition to $p^2 = 0$. In this case, the two tadpoles $I_{\text{Bub}}(1,0)$ and $I_{\text{Bub}}(0,1)$ are equivalent by IBP relations alone, without invoking the symmetry relations. This is the so-called ``magic-relations'', which leads to subtleties in the sector-by-sector dimension counting, since such relations are only evident when considering the three sectors (the top-sector and two sub-sectors) together. In the one-loop bubble case, this relation follows from the fact that in the limit $m_1 = m_2 = m$ and $p^2 = 0$, the $G$ polynomial becomes a function of the sum $x_1 + x_2$, but not of the two variables separately. By a simple variable change $\eta = x_1 + x_2$, the integrals in this family can be written as
\begin{equation}
	\int_0^\infty \dif x_1 \, \dif x_2 \, x_1^{\rho+\nu_1-1} \, x_2^{\rho+\nu_2-1} G^{-d/2} = \int_0^\infty \dif \eta \, G(\eta)^{-d/2} \int_0^\eta \dif x_1 \, x_1^{\rho+\nu_1-1} \, (\eta - x_1)^{\rho+\nu_2-1} ,
\end{equation}
where we have made the regulators explicit to account for the cases when $\nu_i \leq 0$. Through integration by parts, one can lower $\nu_1$ to $0$ while increasing $\nu_2$, which shows that the two sub-sectors are connected by IBP relations. In practice, these relations can always be found by checking the appearance of such situations in all sectors. This can also be validated by counting the dimension with the regularized $u$ function: $u = x_1^\rho \, x_2^\rho \, G^{-d/2}$. For the case at hand, there is only one MI which can be chosen as $I_{\text{Bub}}(1,0)$, i.e., $\hat{e} = \delta(x_2)$. The dual basis vector needs to be taken as $\hat{h} = 1/x_1 + 1/x_2 $, since both the relative boundaries $x_1 = 0$ and $x_2 = 0$ should be accounted for. We have checked that this leads to the correct reduction relations, including $I_{\text{Bub}}(0,1) = I_{\text{Bub}}(1,0)$. Note that similar considerations have been discussed in the Appendix of \cite{Caron-Huot:2021xqj}.

\section{Summary}\label{sec:Summary}

In this paper, we initiate the application of relative cohomology and intersection theory to the reduction of loop integrals in the Feynman parametrization. In our approach, the sub-sector integrals correspond to boundary-supported forms in relative cohomology. We can then treat all sectors in an integral family using the Symanzik polynomials of the top sector, which becomes the twist in the language of intersection theory. With an appropriate choice of the dual basis, the reduction of Feynman integrals can be achieved by computing the relevant intersection numbers.

The usage of the Feynman parametrization has a few advantages over the Baikov representation that is usually employed in the literature. The Symanzik polynomials are homogeneous in the Feynman parameters, and are usually simpler than the Baikov polynomial. In the Feynman parametrization, one does not need to introduce ISPs unless they appear in the numerator. When that happens, the integral can be easily represented by integrands with Symanzik polynomials in the denominator. While these integrands appear to be Feynman integrals in shifted spacetime dimensions, they can be straightforwardly reduced within intersection theory without extra burden. Finally, the dual basis can always be chosen to have at most simple poles at the relative boundaries. All the above lead to the fact that performing integral reduction using intersection theory and relative cohomology in the Feynman parametrization can be simpler than in the Baikov representation.

We have applied our method to several simple examples to demonstrate the correctness of our approach. To improve the efficiency further and to tackle more difficult problems, it will be necessary to employ the technology of finite fields and modular arithmetics~\cite{Peraro:2016wsq, Peraro:2019svx, Klappert:2019emp, Magerya:2022hvj, Mokrov:2023vva, Liu:2023cgs, Mangan:2023eeb}. There are also recent developments in the computation of intersection numbers \cite{Chestnov:2022xsy, Chestnov:2022alh, Brunello:2024tqf}, that can potentially be adopted in our method as well. Finally, in the Baikov representation, one may apply a set of spanning cut to simplify the calculation. In the Feynman parametrization, cutting a propagator corresponds to removing a boundary of the integration domain \cite{Britto:2023rig}. It is therefore well possible to adopt the same trick of spanning cut in our approach. We leave these for future investigation.

\begin{acknowledgments}
The authors would like to thank Hjalte Frellesvig and Xuhang Jiang for useful discussions and comments on the manuscript.
This work was supported in part by the National Natural Science Foundation of China under Grant No. 12375097, 12347103, and the Fundamental Research Funds for the Central Universities.
\end{acknowledgments}

\bibliographystyle{JHEP}
\bibliography{references_inspire.bib,references_local.bib}

\end{document}